# Effect of Rotational Speed on the Stability of Two Rotating Side-by-side Circular Cylinders at Low Reynolds Number

**DOU Hua-Shu[1]\*, ZHANG Shuo[1], YANG Hui[1], SETOGUCHI Toshiaki[2], KINOUE Yoichi[2]**

[1]Key Laboratory of Fluid Transmission Technology of Zhejiang Province, Zhejiang Sci-Tech University, Hangzhou, Zhejiang 310018, China

[2]Department of Mechanical Engineering, Saga University, Honjo-machi, Saga, 840-8502, Japan

Flow around two rotating side-by-side circular cylinders of equal diameter $D$ is numerically studied at the Reynolds number $40 \leq Re \leq 200$ and various rotation rate $\theta_i$. The incoming flow is assumed to be two-dimensional laminar flow. The governing equations are the incompressible Navier-Stokes equations and solved by the finite volume method (FVM). The ratio of the center-to-center spacing to the cylinder diameter is $T/D=2$. The objective of the present work is to investigate the effect of rotational speed and Reynolds number on the stability of the flow. The simulation results are compared with the experimental data and a good agreement is achieved. The stability of the flow is analyzed by using the energy gradient theory, which produces the energy gradient function $K$ to identify the region where the flow is the most prone to be destabilized and the degree of the destabilization. Numerical results reveal that $K$ is the most significant at the separated shear layers of the cylinder pair. With $Re$ increases, the length of the wake is shorter and the vortex shedding generally exhibits a symmetrical distribution for $\theta_i < \theta_{crit}$. It is also shown that the unsteady vortex shedding can be suppressed by rotating the cylinders in the counter-rotating mode.

**Keywords**：numerical simulation, rotating circular cylinders, stability, vortex interactions, energy gradient theory

## Introduction

Flow past a rotating cylinder is one of the most classical problems in fluid mechanics and has been extensively studied numerically and experimentally. The earliest experimental study of the nominally two-dimensional flow past a rotating circular cylinder was conducted by Prandtl in 1925 [1]. Hu et al. [2] found that the rotation may delay the onset of vortex streets and decrease the vortex-shedding frequency. Badr et al. [3] numerically studied the steady and unsteady flows past a cylinder which is rotating at a constant angular velocity and translating with constant linear velocity for $Re$=5, 20, 60, 100 and 200. Kang et al. [4] calculated the flow for $Re$=60, 100, and 160 in the range of $0 \leq \alpha \leq 2.5$. They showed that the value of critical rotation rate $\alpha_{crit}$ increases logarithmically as Reynolds number increases. It is also shown that the Strouhal number is strongly dependent on the Reynolds number. Stojković et al. [5] explored the laminar flow around a circular cylinder with high rotation rates at $Re$=100. The lift force is almost linearly varying with the rotation rate and nearly independent of $Re$ for low rotation rates at $0 \leq \alpha \leq \alpha_{crit}$. They fairly verified the conclusion of the Kang et al. [4]. Vignesh et al. [6] and Prabul et al. [7] performed numerical simulation the effect of non-dimensional rotational rate,

---

*Corresponding author: huashudou@yahoo.com
DOU Hua-Shu: Professor

Reynolds number and Prandtl number of the fluid on laminar forced convection from a rotating horizontal cylinder subject to constant heat flux boundary condition. The rotational effects results in reduction in heat transfer compared to heat transfer from stationary heated cylinder due to thickening of boundary layer.

In recent years, there are a number of studies on the flow around two rotating side-by-side circular cylinders. Yoon et al. [8-9] performed numerical simulation for various absolute rotation rates ($|\alpha_{crit}|\leq2$) for different gap spacing at characteristic Reynolds number of 100, and showed that the critical rotation rates are smaller than that of the single cylinder case. Kumar et al. [10] studied the vortex shedding suppression using PIV at different Reynolds numbers and dimensionless rotation rates for four different spacing ratios for inward and outward rotation configurations. Supradeepan and Roy [11] observed eight different flow regimes for $Re=100$, $\alpha=0$, 0.5, 1.0, 1.25 for $1.1\leq T/D\leq3.5$ using consistent flux reconstruction (CFR) technique. For higher Reynolds number, the flow is too complicated to numerically simulate, the experiment becomes the main approach for investigation at $Re>200$. Zhang et al. [12] further studied the problem of side-by-side rotating cylinders at Reynolds number of 1550 and found that the flow becomes steady and the fluctuation of lift and drag coefficients tend to zero as $|\alpha|$ being beyond the critical rotation rate. Guo et al. [13] analyzed the effects of Reynolds number and rotation on the flow using the particle image velocity (PIV) and a new Immersed-Lattice Boltzmann Method (ILBM) scheme for $425\leq Re\leq1130$, $0\leq\alpha\leq4$ with $T/D=1.11$. They showed the vortex shedding is suppressed as rotation rate increases and the gap spacing has an important role in changing the pattern of vortex shedding.

There are a large number of papers on flow past a single rotating cylinder and two side-by-side counter-rotating cylinders, while most of them focused on the identification of the critical rotation rate and vortex shedding pattern. The limitation of these parameters is that they can only reflect the evolution pattern of the whole flow field but not reveal the specific instability position. In order to clarify these problems, this work investigates the flow around the two counter-rotating cylinders at $40\leq Re\leq200$, $T/D=2$ with different rotation rates and the distributions of several selected quantities are presented. On the other hand, the study on flow instability is also useful in the design of turbomachinery [14]. The physical mechanism of flow instability from the flow around cylinders could provide with reference for performance improvement of turbomachinery.

According to Williamson [15], the wake remains two-dimensional at Reynolds number up to 200. In this paper, the study is focused in the analysis on the control of two-dimensional vortex wakes in the low Reynolds number regime while the effect of three-dimensionality is not considered. The simulation results are compared with those in the literatures. At last, the stability of the flow is analyzed by using the energy gradient theory [16-20].

## Brief Introduction of the Energy Gradient Theory

Dou et al. [16-20] proposed the energy gradient theory to analyze flow instability and turbulent transition, which is based on the Newtonian mechanics and is compatible with Navier-Stokes equations. For a given base flow, the flow stability depends on the relative magnitude of the two roles of energy gradient amplification and viscous friction damping under given disturbance. The flow transits to be unsteady as the ratio beyond a certain critical value. The determination of the criterion of instability can be written as follows:

$$K \frac{v_m'}{u} \leq const \qquad (1)$$

$$K = \frac{V\frac{\partial E}{\partial n}}{V\frac{\partial W}{\partial s}} = \frac{V\frac{\partial E}{\partial n}}{V\frac{\partial E}{\partial s}+\phi} \qquad (2)$$

Here, $K$ is a dimensionless field variable and expresses the ratio of transversal gradient of the total mechanical energy and the rate of the work done by the shear stresses along the streamline; Actually, it is equivalent to a local Reynolds number; $E=p+1/2\rho U^2$ is the total mechanical energy per unit volumetric fluid; $s$ denotes the transversal direction of the streamline, and $n$ is along the normal direction of the streamline. $u$ is the streamwise velocity of main flow; $v_m' = \overline{A}\omega_d$ is the amplitude of the disturbance of velocity, $\overline{A}$ is the amplitude of the disturbance distance, $\omega_d$ is the frequency of disturbance, $\phi$ is an energy dissipation function and is caused by the viscosity and the deformation.

From Eq. (1) and Eq. (2), the critical condition of flow instability and turbulence transition is determined by $K$. The flow in the region with the maximum of $K$, $K_{max}$, firstly loses its stability, and flow area with large value of $K$ will lose its stability earlier than that with small value of $K$.

## Physical Model and Numerical Method

### Geometric Model and Meshing

A pair of circular cylinders of diameter $D=0.02m$ is placed in an unconfined region with uniform incoming free stream flow. The center-to-center distance between the cylinders is $T$. The computational domain extends to $15D$ and 75*D* upstream and downstream of the cylinder, respectively, and the width of the domain is 30*D*. The detailed geometrical schematic of the computational domain is provided in Fig. 1. The block-structured mesh is used for the simulation (see Fig. 2).

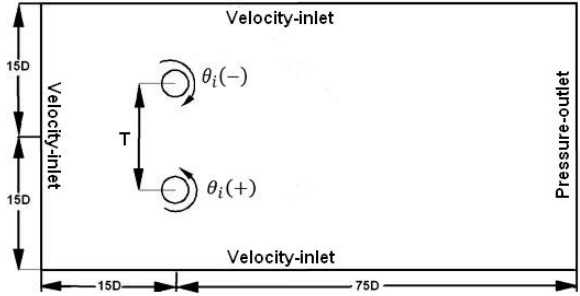

**Fig. 1** Computational domain

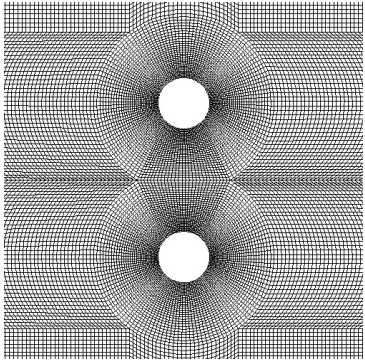

**Fig. 2** Grid around cylinders

### Governing Equations and Numerical Method

The governing equations describing two-dimensional unsteady incompressible laminar isothermal flow in the present study are the continuity and momentum equations which are as follows:

$$\Delta \cdot \vec{v} = 0 \qquad (3)$$

$$\frac{\partial \vec{v}}{\partial t} + (\vec{v} \cdot \nabla) \vec{v} = -\frac{1}{\rho} \nabla p + \upsilon \nabla^2 \vec{v} \qquad (4)$$

Here, $\rho$ is density, $\upsilon$ is the kinematic viscosity, and $\vec{v}$ is velocity vector. Pressure and time are represented by $p$ and $t$, respectively. The discretized equations are solved by the finite volume method (FVM). The coupling of pressure and velocity is performed using the fractional step and non-iterative time advancement (NITA). The time step size $\Delta t$ is given by

$$U \times \Delta t \leq H_{cell} \qquad (5)$$

where, $H_{cell}$ represents the minimum mesh.

### Characteristic Parameter of Flow

The flow pattern is characterized by the following parameters:

(1) Reynolds number $Re = U_\infty D/\upsilon$, $U_\infty$ is the free stream velocity. The Reynolds numbers of $40 \leq Re \leq 200$ are considered in this study.

(2) Non-dimensional rotation rate $\theta_i = \omega D/(2U_\infty)$, $\omega$ is the rotational angular velocity of the cylinder, "±" indicates the direction of rotation, in which counterclockwise is labeled "+".

(3) Center-to-center distance between the cylinders *T*. *T/D*=2 is considered in the present study.

(4) The lift coefficient $C_l = F_l/(1/2 \rho U_\infty^2 D)$ and drag coefficient $C_d = F_d/(1/2 \rho U_\infty^2 D)$, where $F_l$ and $F_d$ represents the lift and the drag force, respectively. The time-average life and drag coefficient are expressed as $\overline{C_l}$ and $\overline{C_d}$.

### Boundary Conditions

As shown in Fig. 1, the uniform free stream condition with constant velocity is imposed at the inlet boundary. At the exit boundary, a pressure outlet boundary condition is defined and the velocity is assumed fully developed. A no-slip boundary condition is prescribed on the cylinder walls. The upper and lower boundary conditions are set to a periodic boundary condition.

### Grid Independence Study

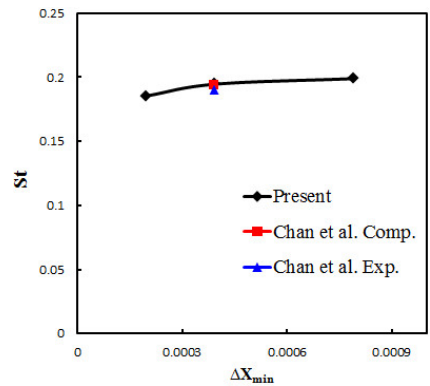

**Fig. 3** Grid independence study

In this paper, the case of *Re*=200, *T/D*=2, $\theta_1$=-1 and $\theta_2$=1 is taken to examine the grid independence of the simulation results. A total grid size of 69240, 277430 and 1126678 control volumes are found to be sufficient to obtain grid independent result. Here, they are marked as $A_1$, $A_2$ and $A_3$, respectively. The finer mesh is essentially reconstructed from the coarser mesh by simply multiplying the number of cells in each direction. The Strouhal number $S_t (= fD/U_\infty)$ is chosen as the reference quantity to validate the simulation results, and is given in Fig. 3 as compared with those in literatures. It can be found from Fig.3 that good agreement is achieved among the results, and there is no noticeable difference between the results obtained by mesh $A_2$ and $A_3$. Finally, the mesh $A_2$ is used in the following simulations.

**Results and Discussions**

Simulations are performed at $Re$=100, 150, 200, $T/D$=2 and various $|\theta_i|$. Fig. 4(a) and (b) show the variations of time-averaged lift and drag coefficients versus $\theta_i$ for the upper cylinder in the counter-rotating, compared with those numerically obtained by Chan et al. [21] and Chan and Jameson [22]. The coefficients for the lower cylinder have the same amplitude but opposite sign for $\overline{C_l}$. For all Reynolds numbers investigated, $\overline{C_l}$ increases linearly with the $\theta_i$ in the range $0 \leq \theta_i \leq 2.6$~2.8, and $\overline{C_l}$ gradually decreases with $\theta_i$ in the range of $0 \leq \theta_i \leq 2.5$~2.6. The magnitudes of the coefficients are in excellent agreement with previous experimental results [21-22].

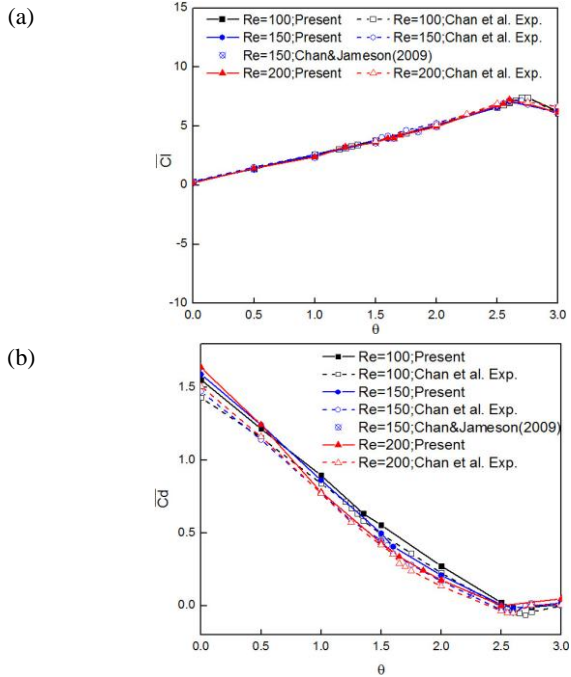

**Fig. 4** Time-averaged (a) lift coefficient ($\overline{C_l}$) and (b) drag coefficient ($\overline{C_d}$) for the counter-rotating cylinder cases ( upper cylinder )

**Effect of the Rotation Rate**

Figure 5 shows the instantaneous distributions of streamwise velocity, streamline, vorticity and $K$ at $Re$=150 for different rotation rates, respectively. As shown in the first column of Fig. 5, the flow with low streamwise velocity is observed in the wake downstream of each cylinder and the gap region. The velocity decreases to almost zero as the rotation rate increases to $|\theta_i|>2.7$. Rotation has a significant effect on the location of the reverse flow zone.

The streamlines and vorticity contours are shown in the second column of Fig. 5. In Fig. 5(a) and (b), the wake shedding switches between the in-phase mode and anti-phase mode as the flow time, in which the gap flow is irregularly moving either upward or downward due to the stronger attraction of upper and lower cylinders shear layer in the gap region, leading to an unstable and asymmetric wake. This enables vortices to develop from the cylinder surfaces. As the rotation rate increases, the interaction of the two shear layers from the two cylinders gets weakened in the gap region. For Fig. 5(c), it can be seen that the vortex formed behind the cylinders oscillates and does not shed from the cylinder surfaces. According to the existing literature [22], the rotation rate for the suppression of the unsteady wake shedding is about $\theta_i$ =1.5~1.6 for this condition. As shown in Fig. 5(d), it exhibits vorticity lobes extend downstream and no sign of shedding as a result of the steady flow. By further increasing the rotational speed, the region with high vorticity magnitude moves closer to the surfaces of cylinder pair, and even fully concentrated on the cylinder surfaces at $|\theta_i|$=3. These streamlines show the presence of the stagnation point in the wake of the cylinder pair, as seen in Fig. 5(e).

The third column of Fig. 5 shows the $K$ contour. For stationary cylinder pairs, the larger value of $K$ is located in the lateral side of each cylinder and downstream wake. This phenomenon was also observed by Dou and Ben [23]. By increasing the rotation rate, it can be seen that the $K$ value distribution tends to be symmetrical. In cases with no wake shedding, $K_{max}$ is observed at the lateral sides of the cylinders and no noticeable evidence is seen in the downstream region as shown Fig. 5(d). According to the energy gradient theory, the location of $K_{max}$ represents that the flow will firstly lose its stability. It indicates that the shear layer of the both sides of cylinders have great effect on flow stability. Moreover, the value of $K$ decreases gradually in the gap region due to the inhibition of rotation.

To explore the relation between the $K$ distribution and rotation rate, the corresponding $K$ profiles downstream of the cylinder pair are shown in Fig. 6. These profiles are taken at a distance of different multiples of diameters downstream of the cylinder pair. The $K_{max}$ appears at the lateral side of cylinder, i.e. the position of the shear layer close to the upper cylinder, as shown Fig. 6(a) and (b), while the larger value of $K$ can still be spotted in the downstream vortices. In Fig. 6(c), $K_{max}$ is observed at the shear layer in the gap region. It can be regarded that instability occurs in the shear layer located in gap region. This is analogous with the conclusion obtained by Supradeepan and Roy [9] that the decisive factor of vortex shedding is the strong interaction between the vortices generated from the cylinder surfaces and the shear layer. There are two locations for the larger value of *K*, which are both at the shear layer in the gap, as shown Fig. 6(d). By further increase the rotation speed, the tangential velocity of the cylinder in the gap is opposite to that of the incoming flow, thus the gap flow is suppressed. The larger value of *K* exists only in the position close to the cylinders surfaces shown in Fig. 5(e). The unsteady wake shedding has been completely suppressed. We can generally conclude here that due to the increase of the rotation rate, the vortex generation from the cylinder surfaces are suppressed, thus the interaction between the shear layer and vortices are weakened.

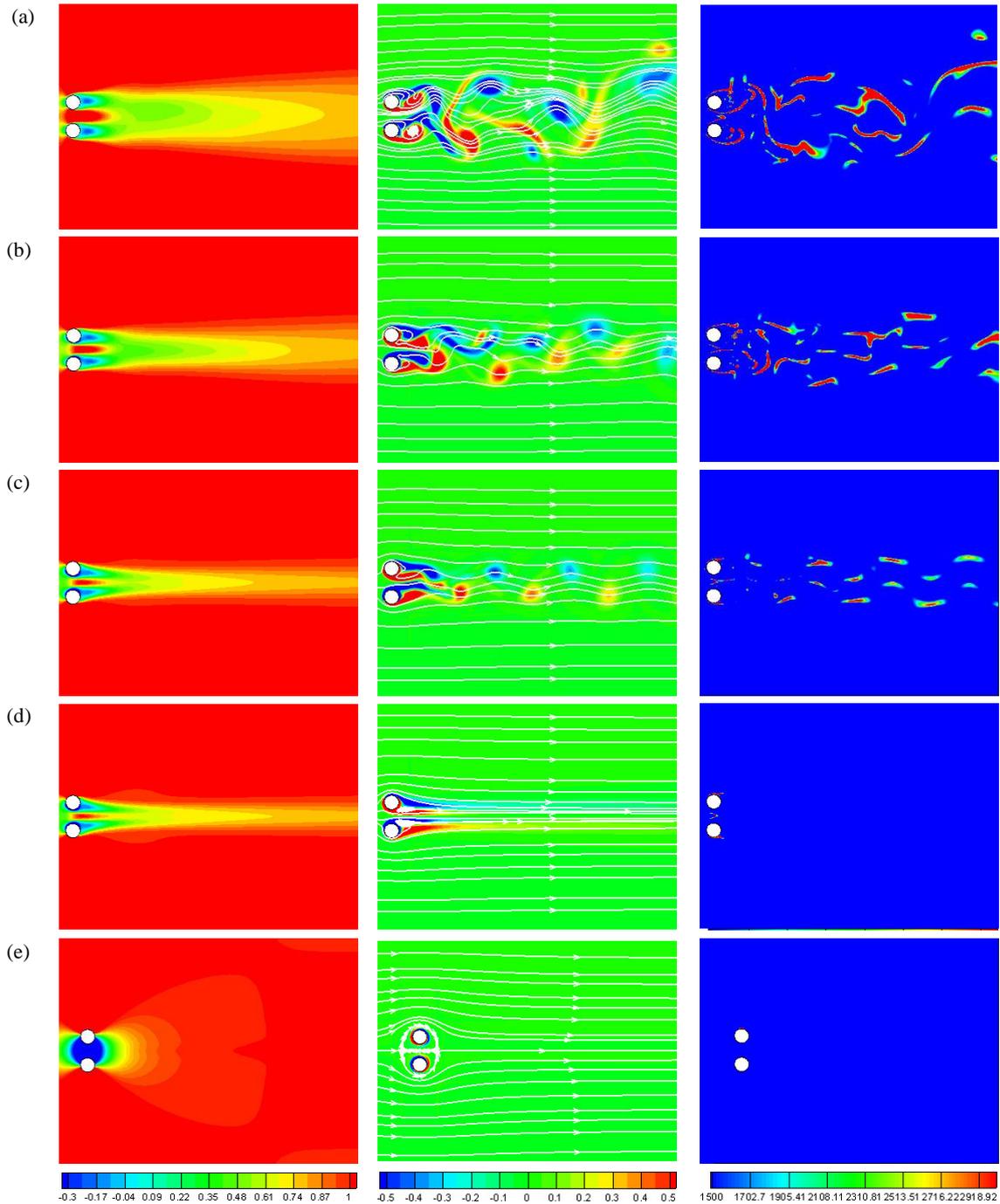

**Fig. 5** Counter-rotating configuration: distribution of dimensionless velocity (first column), streamlines and vorticity (second column) and *K* (third column) for various rotation rates at *Re*=150. (a) $|\theta_i|$=0. (b) $|\theta_i|$=1. (c) $|\theta_i|$=1.5. (d) $|\theta_i|$=1.6. (e) $|\theta_i|$=3.

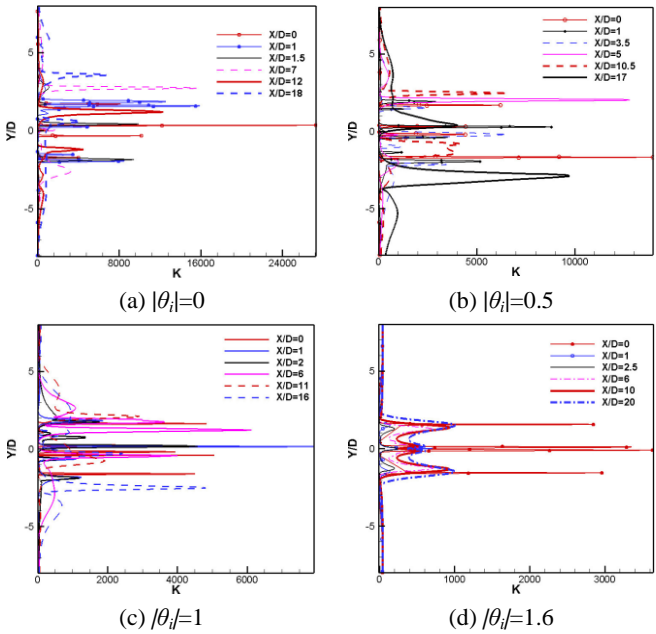

(a) $|\theta_i|=0$  (b) $|\theta_i|=0.5$

(c) $|\theta_i|=1$  (d) $|\theta_i|=1.6$

**Fig. 6** Distribution of $K$ at $Re$=150

### Effect of Reynolds Number

Figures 7 and 8 show the contours of instantaneous velocity, streamline, vorticity and $K$ field at $Re$=40 and 200.

For a pair of stationary cylinders in Fig. 7(a), the flow is steady and symmetric about the centerline without generating vortex shedding, and $K$ is negligible. According to Williamson [15], the flow instabilities occur in the wake of a pair of stationary cylinders with $T/D$=2 for Reynolds number above 55. Increasing the rotation rate, the reverse flow region behind the cylinders diminishes and symmetric vortices detach from cylinder surfaces. Vorticity lobes get shorter due to the influence of the gap flow and have a tendency to concentrate on the surfaces shown in Fig. 7(b) and (c). For the $K$ counter in Fig. 7(b), it can be found that the peak value of $K$ is larger than the case of a stationary cylinder pair. For the $K$ counter in Fig. 7(c), it can be found that $K$ is negligible around the cylinders and the peak value of $K$ is smaller than that of the case of $|\theta_i|$=1.5. It can also be found that the flow patterns are similar for cases of $Re$=40 and 150 at the same $|\theta_i|$=3.

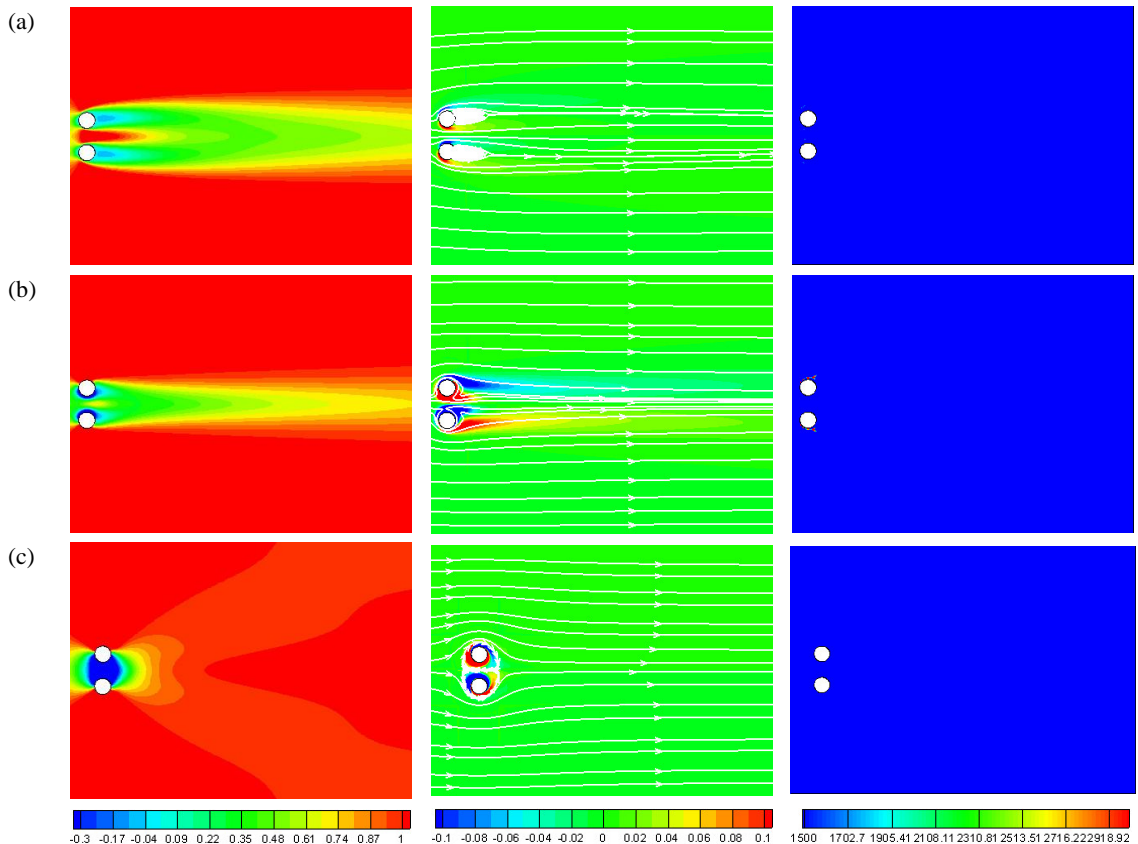

**Fig. 7** Counter-rotating configuration: distribution of dimensionless velocity (first column), streamline and vorticity (second column) and $K$ (third column) for various rotation rates at $Re$=40. (a) $|\theta_i|$=0. (b) $|\theta_i|$=1.5. (c) $|\theta_i|$=3.

In Fig. 8 (a)-(c), the wake shedding mode transits from disordered state to ordered state with increasing rotation rates, which is consistent with the previous results. For $|\theta_i|$=1.5, the wake pattern is characterized by the synchronized vortices from the upper and lower cylinders. Two separate vortex streets form downstream of the cylinder pair without interaction, which indicates that the reduction of the interference between the cylinders. For the case of critical rotation rate 1.65 based on historical data, $K$ is noticeable at the lateral sides the cylinder pair and the .

outer boundaries of the vortex, as shown in Fig. 8(b). For the rotation rate up to 3, the flow patterns are similar to those of $Re$=40 and 150 at the same $|\theta_i|$ as shown in Fig. 5(e), Fig. 7(c) and Fig. 8(c).

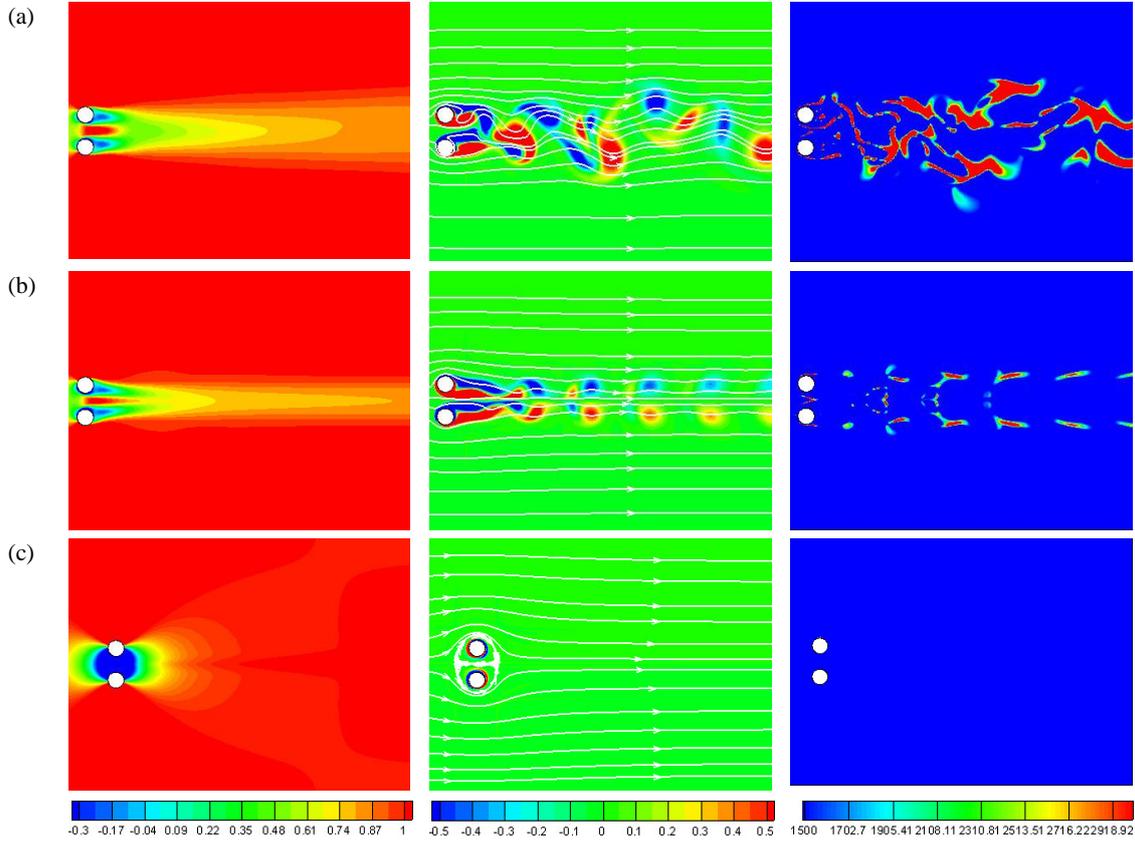

**Fig. 8** Counter-rotating configuration: distribution of dimensionless velocity (first column), streamlines and vorticity (second column) and $K$ (third column) for various rotation rates at $Re$=200. (a) $|\theta_i|$=1. (b) $|\theta_i|$=1.5. (c) $|\theta_i|$=3.

To explore the physical mechanism of the relatively large $K$ value, the corresponding velocity, vorticity and $K$ profiles in the region downstream of the cylinder pair are shown in Fig. 9. It can be seen from the first column of Fig. 9 that the mean streamwise velocity shows similar distributions. Due to the existence of the shear layer, the mean streamwise velocity reaches a maximum value then rapidly reduces to 0 at the cylinder surfaces. The mean streamwise velocity decreases continuously from the vortex edge to the centerline. The distribution of velocity basically keeps unchanged in the external vortices. The mean streamwise velocity in the wake increases as the flow develops downstream. In the third column of Fig. 9(a), $K_{max}$ occurs at the position of inflection point of vorticity and the maximum value of velocity in each profile. It also can be found that the value of $K$ at the lateral sides of two vortices for each cylinder is large, while the value of $K$ at the position with two vortices is low. $K$ decreases significantly in the downstream. This confirms the conclusion that these vortices at the rear of the cylinder have less effect on the flow stability. For Fig. 9(b), $K_{max}$ only locates in the shear layers of the top of upper cylinder and the bottom of the lower cylinder, and the amplitude of $K$ is lower in the gap region. The value of $K$ for each profile is larger than the case of $Re$=40. For the case $Re$=200 shown in Fig. 9(c), the $K_{max}$ exists in the shedding wake vortices and larger value of $K$ is also observed in the shear layer. It can be regarded that the shedding is originated from the combination of the interaction of shear layers for the two cylinders.

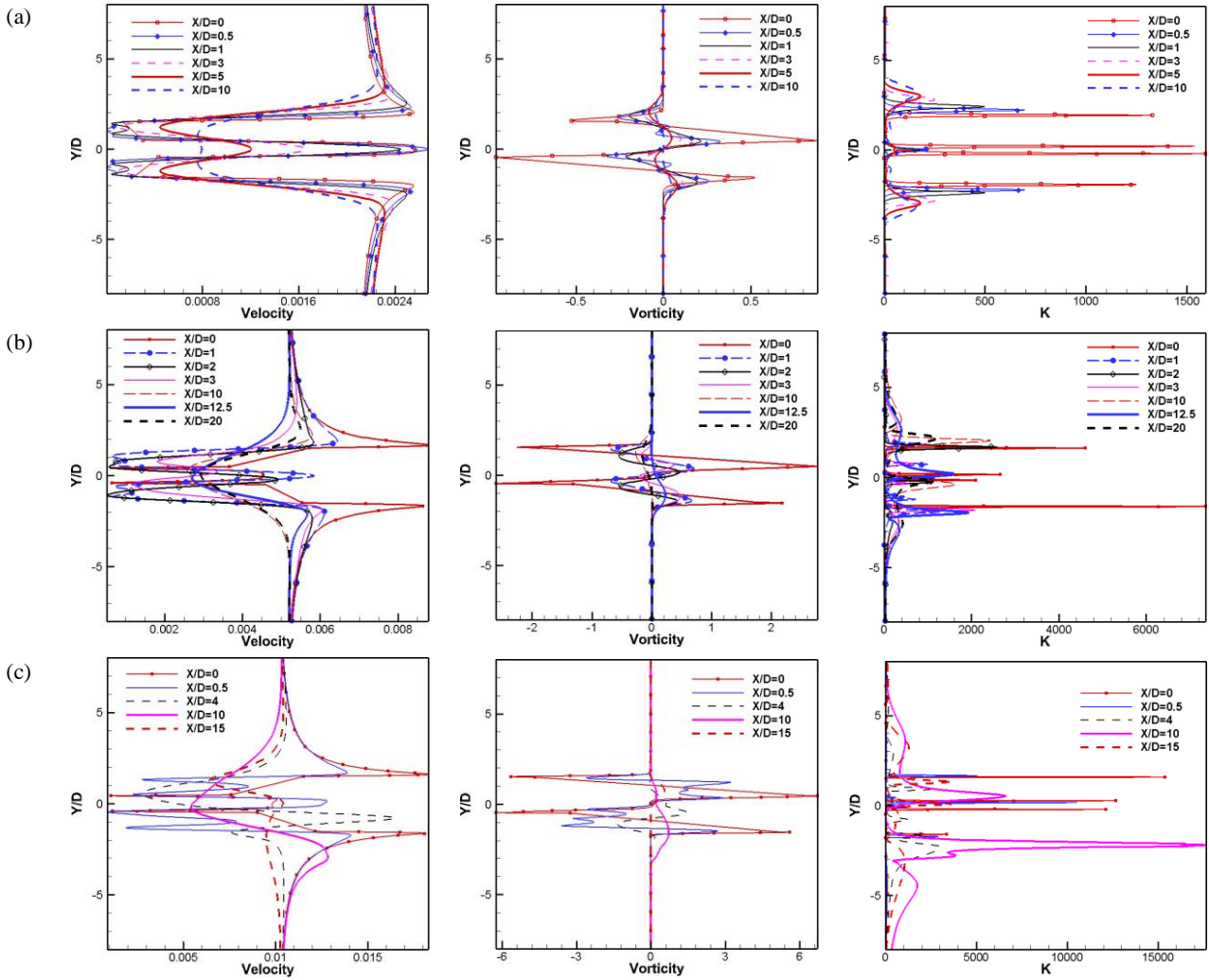

**Fig. 9** Distribution of velocity, vorticity and $K$. (a) $Re$=40, $|\theta_i|$=0. (b) $Re$=100, $|\theta_i|$=1. (c) $Re$=200, $|\theta_i|$=1.

The streamwise distributions of $K$ for $Y/D$=0, 1 and -1 at $|\theta_i|$=1 are shown in Fig. 10. The fluctuation of $K$ decays downstream much slower for $Re$=200 than for $Re$=40, 100 and 150 cases, hence the whole flow field is more unstable. For Fig. 10(a) and (b), a larger value of $K$ is found in the downstream region for $Re$=200, which represents that the instability here will spread towards upstream and affect the stability of the shear layers which are located at lateral sides of cylinder pairs vortices and are with large $K$.

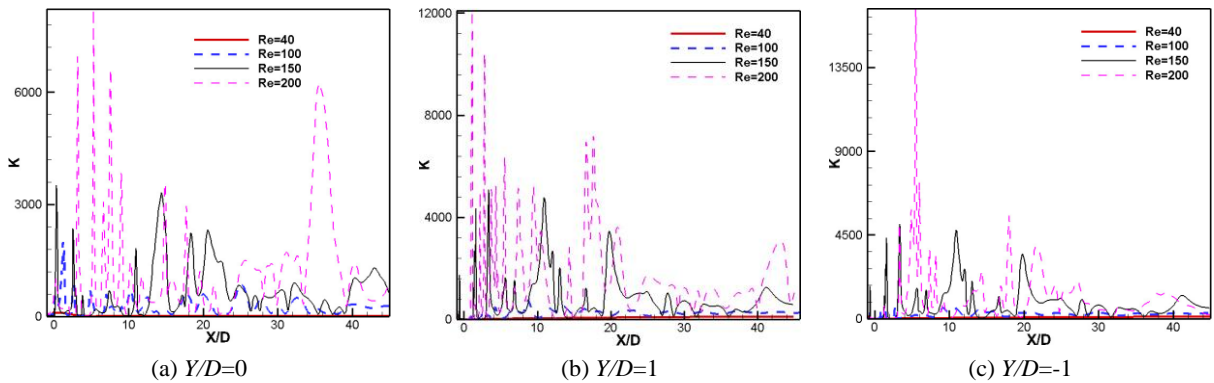

(a) $Y/D$=0        (b) $Y/D$=1        (c) $Y/D$=-1

**Fig. 10** Streamwise distribution of $K$ at $|\theta_i|$=1.

- 8 -

## Conclusions

In summary, the flow past a pair of counter-rotating circular cylinders is computed at different rotation rates for $40 \leq Re \leq 200$ and spacing of 2. The computed results are compared with those in literature and good agreements are obtained. The instabilities of flow field are investigated by using the energy gradient theory. The following conclusions are obtained:

1. The rotation rate has a significant effect on the vortex shedding and flow state in the wake flow at low Reynolds number. The vortex shedding is suppressed as the rotation rate increases. The value of $K_{max}$ gradually decreases as the rotation rate increases. This indicates that the flow instability is weakened with increasing rotation rate.

2. For low rotation rate, the $K$ is mainly distributed at both sides of the cylinder pair shear layer and in the shedding wake vortices. For high rotation rate, $K$ is concentrated in the shear layer of the cylinder at both sides and is low in the shedding wake vortices. Increasing the rotation rate weakens the interference between the shear layer and vortex.

3. The flow patterns are similar for various Reynolds numbers as the rotation rate increases. At the same gap spacing and rotation rate, the value of $K$ increases with the increasing of Reynolds number.

4. The energy gradient theory is potentially applicable to study the flow stability and have a good predictability of instability position.

## Acknowledgement

This work is supported by National Natural Science Foundation of China (51579224), and Zhejiang Province Science and Technology Plan Project (2017C34007).